\documentclass[twocolumn,english,superscriptaddress,floatfix,longbibliography]{revtex4-2}
\usepackage[T1]{fontenc}
\usepackage[latin9]{inputenc}
\usepackage{amsmath}
\usepackage{graphicx}
\usepackage{babel}
\begin{document}
\title{Symmetry Breaking and Spatiotemporal Pattern Formation in Photonic
Time Crystals}
\author{Egor I. Kiselev}
\affiliation{Physics Department, Technion, 320003 Haifa, Israel}
\affiliation{The Helen Diller Quantum Center, Technion, Haifa 3200003, Israel }
\email{kiselev.physics@gmail.com}

\author{Yiming Pan}
\affiliation{School of Physical Science and Technology and Center for Transformative
Science, ShanghaiTech University, Shanghai 200031, China}
\begin{abstract}
In this work, we explore the dynamics of time varying photonic media
with an optical Kerr nonlinearity and an associated phase transition.
The interplay between a periodically modulated permittivity and the
nonlinearity induces a continuous transition of electromagnetic waves
to a state with broken spatial and time translation symmetries. This
transition gives rise to a lattice-like wave pattern, in many ways
similar to a spatial crystallization in solids. Symmetry breaking
triggers the emergence of soft, Goldstone-like modes, which propagate
as deformations of the lattice structure, as well as massive Higgs-like
modes -- spatially uniform oscillations of the field amplitude. We
extend the analysis of the non-equlibrium symmetry breaking to 2+1
dimensional time varying media and discuss pattern formation as well
as the connection to discrete dissipative time crystals.
\end{abstract}
\maketitle

\section{Introduction}

The propagation of electromagnetic fields through media with time
varying material parameters has been considered early on \citep{morgenthaler1958earlyPTC,holberg1966parametric,felsen1970wave_propagation_time_varying}.
More recently, time varying media have attracted attention in photonics
\citep{galiffi2022photonics_of_time_varying}, metamaterials \citep{wang2023metamaterial_ptc,koutserimpas2023multiharmonic_coupled_time_modulated},
and material science \citep{saha2023_ptc_material_perspective}, particularly
in the field of epsilon near zero materials \citep{zhou2020_time_refraction_epsilon_near_zero,bruno2020broad_freq_shift_epsilon_near_zero,pang2021adiabatic_freq_conversion_epsilon_near_zero,tirole2022_time_reversal_mirror}.
They are viewed as promising candidates for realizing such effects
as momentum-gapped (k-gapped) states, nonreciprocity \citep{ramaccia2018nonreciprocity},
time switching \citep{akbarzadeh2018_temporal_switch}, time-varying
mirrors \citep{tirole2022_time_reversal_mirror}, lasing and amplification
\citep{lyubarov2022_ptc_amplified_emission_lasing}. From a fundamental
perspective, the non-trivial statistical properties of randomly driven
time-varying media \citep{carminati2021_random_driven_time_varying_medium_statistics},
their topological \citep{lustig2018topological_aspects_of_ptc} and
radiative \citep{dikopoltsev2022light_emission_free_electrons_ptc,li2023_ptc_light_emission_by_stationary_charge}
properties have been explored. Beyond optics, hydrodynamic \citep{bacot2016time_reversal_hydro,bacot2019hydro_phase_conjugate_mirror}
and acoustic \citep{zhu2023time_varying_acoustic} time varying systems
have been studied. 

In this paper, we study nonlinear photonic time crystals (nonlinear
PTCs) -- dielectric media with a periodically modulated permittivity
\citep{lustig2023ptc_review,asgari2024PTC_review}. It is crucial
to distinguish PTCs from ``time crystals'' and ``discrete time
crystals'' in the sense of Wilczek \citep{wilczek2012quantum_time_crystals,shapere2012classical_time_crystals,zaletel2023colloquium_time_crystals},
which have been intensively studied in theory \citep{heo2010_old_time_crystal2,watanabe2015absence_of_time_crystals}
and experiment \citep{kim2006_old_time_crystal1,zhang2017observation_dtc1,choi2017observation_dtc2,rovny2018observation_dtc3,pal2018temporal_order_obeservation_dtc4,kessler2021observation_ddtc5,taheri2022_observation_ddtc6}.
While it is important to maintain this distinction, we demonstrate
that, in the presence of a Kerr term, a PTC evolves to a subharmonic
steady state oscillating at half the driving frequency, and thus breaks
the discrete time translation symmetry of the drive on long timescales.
The steady state is very much akin to Faraday waves \citep{Faraday1837_faraday_waves,benjamin1954_faraday_waves,edwards1994_faraday_waves},
which are in many ways similar to classical discrete time crystals
\citep{yao2017discrete_classical_time_crystals,else2020discrete_time_crystals,zaletel2023colloquium_time_crystals}.
In fact, an observation of Faraday waves in a cold atom Bose-Einstein-Condensate
(BEC) \citep{engels2007observation_faraday_cold_atoms} has recently
been reinterpreted as experimental evidence for a discrete quantum
space-time crystal \citep{smits2018observation_space_time_crystal_cold_atoms,liao2019_theory1_space_time_crystal_cold_atoms,smits2020_theory_exp_3_space_time_crystal_cold_atoms,smits2021_theory2_space_time_crystal_cold_atoms,stehouwer2021_theory4_space_time_crystal_cold_atoms}.
An axial mode of a BEC trapped in a cigar-like potential was parametrically
excited by a long-lived, radial oscillatory mode. This led to the
formation of a periodic spatial structure in the axial direction.

One of the most striking features of PTCs is the so called momentum
gap (sometimes referred to as k-gap or q-gap, see Fig. \ref{fig:Fig_transition}a)
-- an interval of wavenumbers in which the real part of the dispersion
$\omega\left(q\right)$ becomes flat and the imaginary part changes
sign, leading to exponentially growing modes inside the gap \citep{lustig2023ptc_review}.
In nonlinear PTCs and time varying media, the k-gap can lead to intriguing
effects such as the formation of superluminal solitons \citep{basov1966_superluminal_pulse,sazonov2001superluminal_solitons,pan2023superluminal_solitons_PTC,cohen2023annihilation_superluminal_solitons}.
We show that in PTCs featuring a generic Kerr nonlinearity, this growth
heralds an instability, leading to the emergence of non-equilibrium,
symmetry breaking steady-states: When the PTC's permittivity is modulated
at a frequency $2\Omega$, the field adopts a standing wave lattice
pattern whose wavenumber $q^{*}$ and lattice constant $\lambda=2\pi/q^{*}$
are determined by the condition for parametric resonance $q^{*}=\Omega/c$
(Fig. \ref{fig:Fig_transition}).

Furthermore, we show that in the symmetry breaking state, slow, long-wavelength
electromagnetic fields propagate through the PTC as distortions of
the lattice, locally contracting or expanding the lattice constant
$\lambda$ (see Fig. \ref{fig:Fig_collective_modes}). This behavior
is reminiscent of phonons in a crystal lattice, and the corresponding
distortions can be thought of as the Goldstone modes of the symmetry
broken state. Additionally, the symmetry broken state hosts spatially
uniform oscillations of the field amplitude at a characteristic frequency
that depends on the driving strength (see Fig. \ref{fig:Fig_collective_modes}b).
These oscillations are gapped, massive modes that resemble, e.g.,
the Higgs modes of a superconductor.

PTCs have already been realized as metamaterials in the GHz range
\citep{wang2023metamaterial_ptc}. Plasmonic realizations that are
within experimental reach and could extend the frequency range to
THz have been proposed \citep{wilson2018_plasmon_time_varying_1,kiselev2023light,kiselev2023inducing,shirokova2023_plasmon_time_varying_2}.
A promising route to realizing PTCs at optical frequencies are epsilon
near zero materials \citep{saha2023_ptc_material_perspective} where
the time refraction of optical signals has recently been demonstrated
\citep{lustig2023time_refraction_epsilon_near_zero} -- a promising
milestone on the way to creating PTCs at optical frequencies.

\section{Results}

\subsection*{Symmetry breaking transition.}

\begin{figure}
\centering{}\includegraphics[width=1\columnwidth]{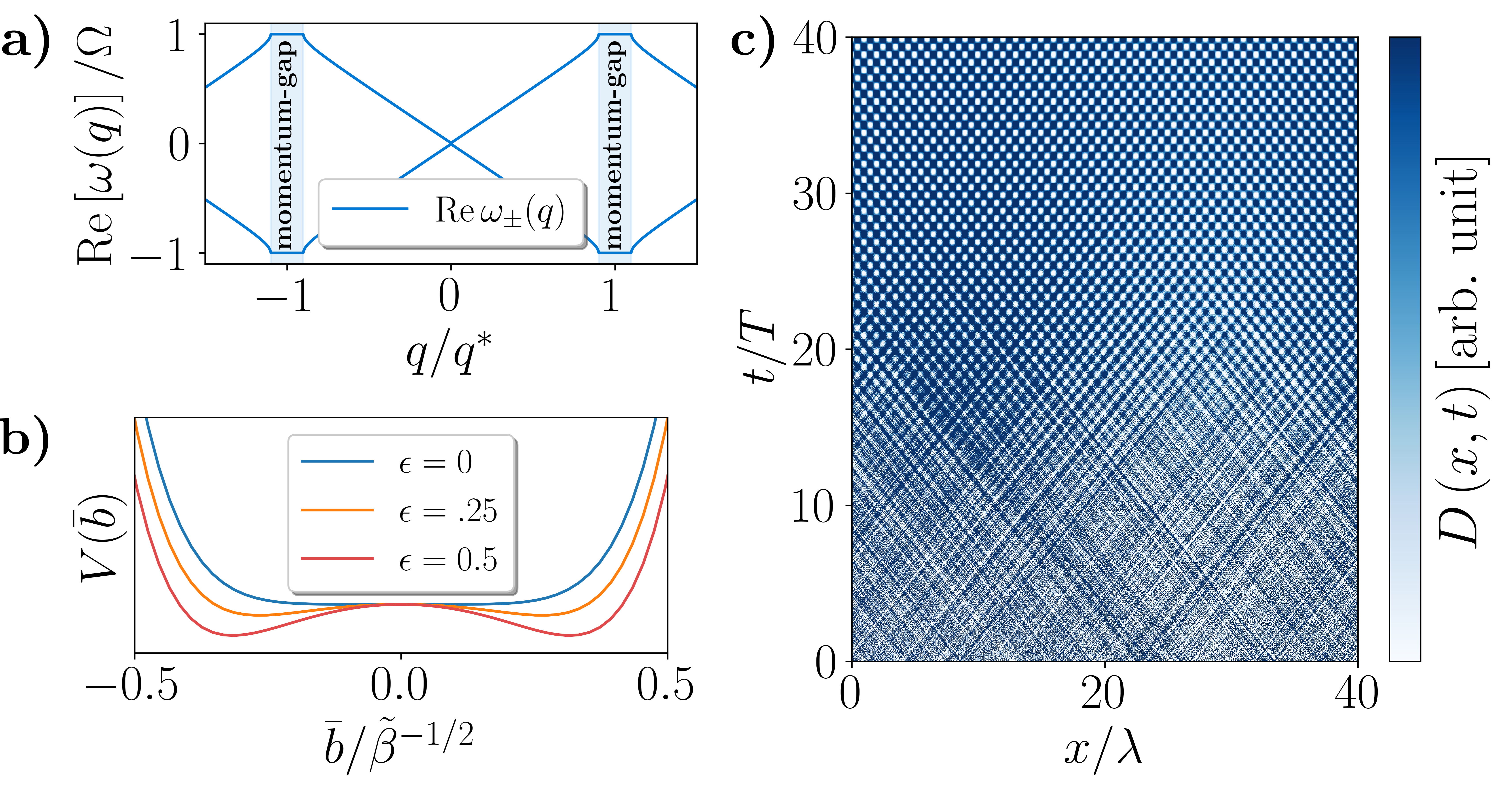}\caption{\textbf{a)} Real part of the dispersion relation $\omega\left(q\right)$
of the PTC with momentum gaps at $q^{*}=\Omega/c$, where $2\Omega$
is the modulation frequency of the dielectric constant. \textbf{b)}
Effective potential $V\left(\bar{b}\right)$ for the displacement
field amplitude $\bar{b}$ with minima corresponding to steady states
with broken translational and discrete time translational symmetries
(see Eq. (\ref{eq:symmetry_breaking_potential})). Here, $\epsilon$
is a parameter that encodes the relative strength of the modulation
of the dielectric constant with respect to damping (see Eq. (\ref{eq:epsilon_def})).
\textbf{c)} A numerical solution of Eq. (\ref{eq:PTC_eq}). Starting
from a random field distribution at $t=0$, the plot shows the spatio-temporal
evolution of the field $D\left(x,t\right)$ to the symmetry breaking
state (checkerboard-like pattern) given in Eq. (\ref{eq:small_amp_ansatz})
and corresponding to a minimum of the potential $V\left(\bar{b}\right)$.
To produce the plot, we choose a modulation strength close to the
threshold yielding $\epsilon\approx0.02$. The $t$- and $x$-axes
are drawn in units of $T=2\pi/\Omega$ and $\lambda=2\pi/q^{*}$.\label{fig:Fig_transition} }
\end{figure}

In this paragraph, we investigate the symmetry breaking transition
of nonlinear PTCs. We show that the transition is continuous when
the modulation strength of the dielectric constant exceeds a threshold
value determined by the damping.

The propagation of electromagnetic fields in a nonmagnetic, isotropic,
one-dimensional PTC with a generic, repulsive Kerr-nonlinearity is
described by the equation \citep{pan2023superluminal_solitons_PTC}
\begin{equation}
\frac{1}{c^{2}}\frac{\partial^{2}D}{\partial t^{2}}+\frac{\gamma}{c^{2}}\frac{\partial D}{\partial t}-\left(1+h\cos2\Omega t\right)\frac{\partial^{2}D}{\partial x^{2}}=\beta\left|D\right|^{2}\frac{\partial^{2}D}{\partial x^{2}},\label{eq:PTC_eq}
\end{equation}
where $D\left(x,t\right)$ is the displacement field, $\gamma$ is
the decay rate of the electromagnetic field, $c$ is the speed of
light and $h$ is a dimensionless parameter controlling the modulation
strength of the dielectric constant. In what follows we assume $h>0$.
For brevity, we omit vector signs. For a small $h\ll1$, an approximate
solution to Eq. (\ref{eq:PTC_eq}) can be found with the slowly varying
envelope approximation. From the condition of parametric resonance
\citep{Landau_Lifshitz_Mechanics}, we know that an oscillating system
is parametrically excited when the modulation frequency is equal to
twice the resonance frequency divided by an integer $n$. The $n=1$
resonance is the most pronounced. Correspondingly, we anticipate that
the time dependent term will lead to a parametric response at half
the modulation frequency $\Omega$ \citep{Landau_Lifshitz_Mechanics},
and we choose the ansatz
\begin{equation}
D=a\left(t\right)\cos\left(\Omega t\right)\cos\left(q^{*}x\right)+b\left(t\right)\sin\left(\Omega t\right)\cos\left(q^{*}x\right).\label{eq:small_amp_ansatz}
\end{equation}
Here $q^{*}$ is a critical wavenumber determined by the condition
\begin{equation}
\Omega=c\left|q^{*}\right|.\label{eq:resonance condition}
\end{equation}

To first order in $h$, and projecting the nonlinear term of Eq. (\ref{eq:PTC_eq})
onto the ansatz of Eq. (\ref{eq:small_amp_ansatz}), i.e. disregarding
harmonics oscillating at frequencies $\pm3\Omega$ and faster, we
obtain 
\begin{align}
\dot{a} & =-\frac{\gamma}{2}a-\tilde{h}\Omega b+\tilde{\beta}\Omega b\left(a^{2}+b^{2}\right)\nonumber \\
\dot{b} & =-\frac{\gamma}{2}b-\tilde{h}\Omega a-\tilde{\beta}\Omega a\left(a^{2}+b^{2}\right),\label{eq:ampl_eoms}
\end{align}
with $\tilde{h}=h/4$ and $\tilde{\beta}=9\beta/32$.

We deduce from the amplitude equations (\ref{eq:ampl_eoms}) that
for $h<h_{c}$, where 
\begin{equation}
h_{c}=2\gamma/\Omega,\label{eq:h_c_def}
\end{equation}
the PTC is stable around the trivial fixed point $D\left(x,t\right)=0$.
For $h>h_{c}$ this fixed point becomes unstable, leading an initially
exponential growth of the amplitudes $a$, $b$. The nonlinear terms
then become increasingly important, and force the amplitudes to saturate.
Eventually, the PTC reaches a new fixed point, which corresponds to
the symmetry breaking phase that we want to study. 

It is useful to consider two distinct regimes. In the first regime,
which we will call this the \textit{near-critical regime}, $h$ is
close to the critical threshold value $h_{c}$. In this case, both,
the transition speed and the amplitudes $a$, $b$ at the fixed point
are determined by the small parameter
\begin{equation}
\epsilon=\frac{h-h_{c}}{h_{c}}.\label{eq:epsilon_def}
\end{equation}
In the second regime, $h\gg h_{c}$ holds. Here, the damping can be
disregarded to a good approximation. Let us term this the \textit{weakly-damped
regime}. With $\gamma$ set to zero, the equations (\ref{eq:ampl_eoms})
can be derived from the Hamiltonian 
\begin{equation}
H\left(a,b\right)=\frac{\tilde{h}\Omega}{2}\left(a^{2}-b^{2}\right)+\frac{\tilde{\beta}\Omega}{4}\left(a^{4}+b^{4}\right)+\frac{\tilde{\beta}\Omega}{2}a^{2}b^{2}.\label{eq:a_b_hamiltonian}
\end{equation}
Formally, we can think of $a$ as a momentum variable and $b$ as
a coordinate. The phase profile corresponding to $H\left(a,b\right)$
and the trajectories of $a$ and $b$ during the symmetry breaking
phase transition are shown in Fig. \ref{fig:Fig_collective_modes}c.
The minima of the Hamiltonian (\ref{eq:a_b_hamiltonian}) are the
stable fixed points of Eqs. (\ref{eq:ampl_eoms}), and are located
at
\begin{equation}
a_{0}=0,\ b_{0}=\pm\sqrt{\frac{\tilde{h}}{\tilde{\beta}}}.\label{eq:a_b_minima}
\end{equation}
For a small damping $\gamma\ll h\Omega$, the solutions (\ref{eq:a_b_minima})
are slightly modified. To first order in $\gamma/h\Omega$, we find
$a_{0}\approx\mp\frac{\gamma}{\Omega\sqrt{8\tilde{\beta}\tilde{h}}}$
and $b_{0}\approx\pm\sqrt{\frac{\tilde{h}}{\tilde{\beta}}}\mp\frac{\gamma^{2}}{16\Omega^{2}\sqrt{\tilde{\beta}\tilde{h}}}$.
For the most part (except when studying the transition to the symmetry
breaking state), we will not be interested in the damping, and only
include it for the sake of numerical stability. For weak damping,
the system will evolve to a state described by the above solution,
and the $D$-field will be given by
\begin{equation}
D_{0}\left(x,t\right)=a_{0}\cos\left(\Omega t\right)\cos\left(q^{*}x\right)+b_{0}\sin\left(\Omega t\right)\cos\left(q^{*}x\right).\label{eq:sol_1D}
\end{equation}
This steady state of the nonlinear PTC breaks the continuous spatial
translation, and the discrete time translation symmetries of Eq. (\ref{eq:PTC_eq}).
The latter follows from the fact that the solution $D_{0}\left(x,t\right)$
of Eq. (\ref{eq:sol_1D}) is, unlike Eq. (\ref{eq:PTC_eq}) not invariant
under the time translation $t\rightarrow t+\frac{\pi}{\Omega}$.

Let us now turn to the weakly damped regime where the modulation strength
$h$ is close to the instability threshold $h_{c}$. This regime is
useful to study the transition dynamics from the trivial to the symmetry
breaking state. In particular, it can be shown that the transition
into the symmetry breaking steady state is a continuous phase transition.
To this purpose, we rewrite the amplitude equations for $a\left(t\right)$,
$b\left(t\right)$ in terms of the new variables $\bar{a}\left(t\right)=a\left(t\right)+b\left(t\right)$,
$\bar{b}\left(t\right)=a\left(t\right)-b\left(t\right)$. A perturbative
expansion of Eqs. (\ref{eq:ampl_eoms}) in the small parameter $\epsilon$
of Eq. (\ref{eq:epsilon_def}) (see supplement) then predicts that
the transitional dynamics can be described by a gradient descent equation
for $\bar{b}$$\left(t\right)$:
\[
\dot{\bar{b}}=-\frac{\partial V\left(\bar{b}\right)}{\partial\bar{b}},
\]
where $V\left(\bar{b}\right)$ is an effective double-well potential:
\begin{equation}
V\left(\bar{b}\right)=-\epsilon\frac{\tilde{h}_{c}\Omega}{2}\bar{b}^{2}+\frac{\tilde{\beta}^{2}\Omega}{48\tilde{h_{c}}}\bar{b}^{6}.\label{eq:symmetry_breaking_potential}
\end{equation}
We plot the effective potential for different $\epsilon$ in Fig.
\ref{fig:Fig_transition}b and conclude that the transition into the
symmetry breaking state is continuous, i.e. the system goes through
a soft bifurcation when $h$ reaches the critical driving strength
$h_{c}$, which is determined by the ratio of $\gamma$ and $\Omega$
(see Eq. (\ref{eq:h_c_def})). In terms of the original $a$ and $b$,
the potential minima (and approximate fixed points of Eqs. (\ref{eq:ampl_eoms}))
are located at
\begin{equation}
a_{0}=-b_{0}=\pm\sqrt{\frac{\tilde{h}_{c}}{\tilde{\beta}}}\left(\frac{\epsilon}{2}\right)^{1/4}.\label{eq:close_threshold_fixed_points}
\end{equation}
The result of a numerical simulation of Eq. (\ref{eq:PTC_eq}) for
$\epsilon=0.005$ is depicted in Fig. \ref{fig:Fig_transition}c,
showing how the displacement field inside the PTC begins to self organize
itself into the standing wave pattern predicted in Eqs. (\ref{eq:small_amp_ansatz})
and (\ref{eq:close_threshold_fixed_points}). All simulations in this
paper were carried out using the Dedalus software package \citep{burns2020dedalus}.

\subsection*{Emergent collective modes.}

In the previous section, we demonstrated that the unstable, exponentially
growing modes of a nonlinear PTC herald the onset of a new phase.
After a short transitional period, the nonlinear PTC enters a steady
state with broken spatial translation and discrete time translation
symmetries. In this section, we focus on small fluctuation around
this new state. We demonstrate that the lattice structure of the electric
field supports soft, wavelike excitations, which propagate through
the lattice, similar to phonons in a crystal lattice. These excitations
are Goldstone-like modes that stem from the breaking of the continuous
spatial translation symmetry. Additionally, the steady state exhibits
a gapped mode, which is absent in the symmetric state.
\begin{figure}
\centering{}\includegraphics[width=1\columnwidth]{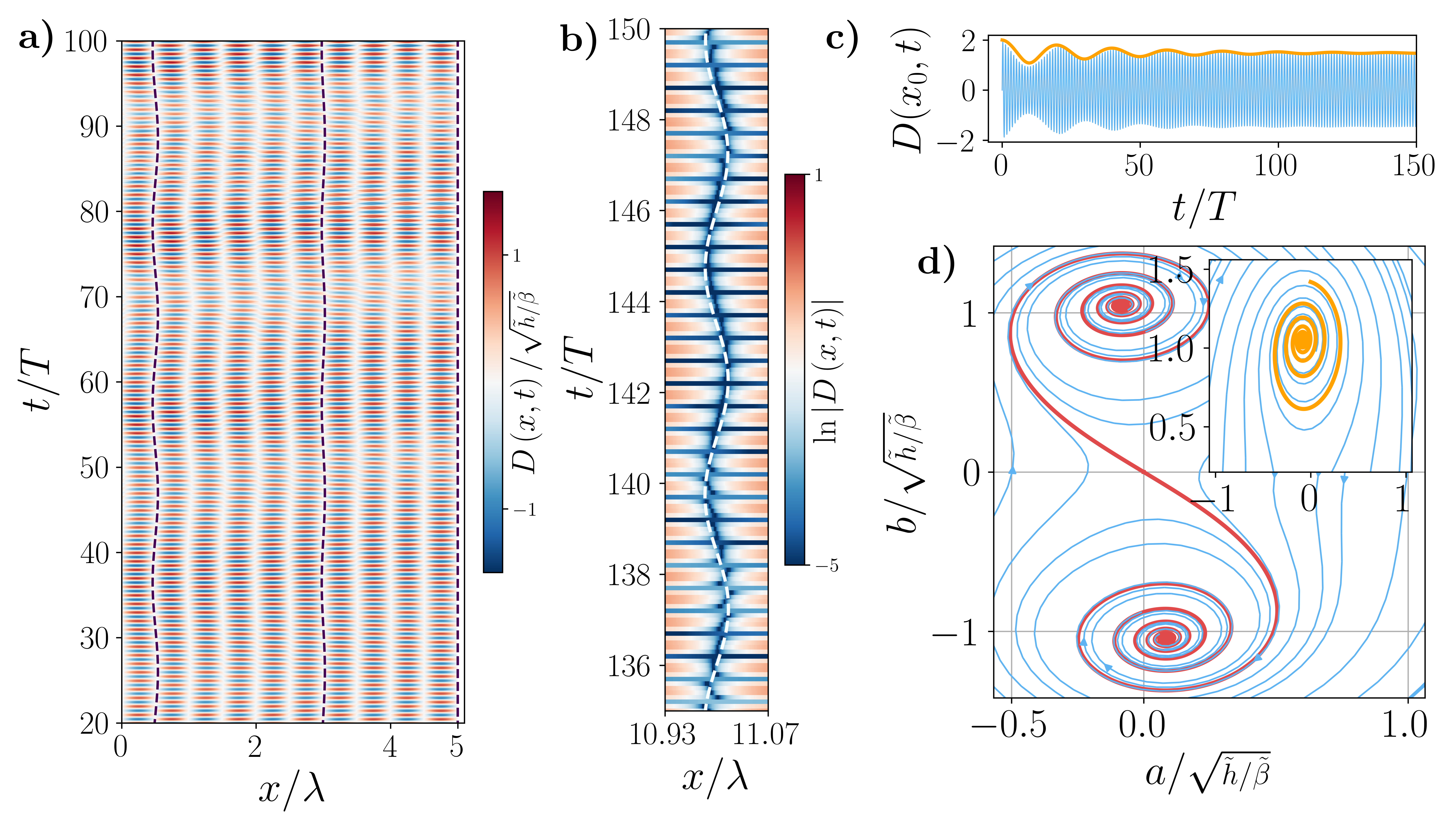}\caption{\textbf{a)} Goldstone-like distortions propagating through the electric
field lattice of the PTC in the symmetry broken state. The displacement
field is given by $D\left(x,t\right)=\sqrt{\tilde{h}/\tilde{\beta}}\sin\left(\Omega t\right)\cos\left(q^{*}x+\phi\left(x,t\right)\right)$.
An oscillatory boundary condition of the form $\phi\left(x=0,t\right)=-0.3\sin\left(\omega_{G}\left(Q\right)t\right)$
is applied. The propagation of the phase $\phi\left(x,t\right)$ through
the wave-lattice is described by Eq. (\ref{eq:goldstone_eq}). The
dashed black curves show the nodes of the standing wave D-field, described
by $q^{*}x+\phi\left(x,t\right)=\pi/2$. \textbf{b) }A Goldstone-like
mode as observed starting with an intitial condition $D\left(x\right)=\sqrt{\tilde{h}/\tilde{\beta}}\cos\left(q^{*}x-0.2\sin\left(q^{*}x/5\right)\right)$
at $t=0$. The Goldstone-like mode propagates according to Eq. (\ref{eq:goldstone_eq})
(white dashed line), where small damping $\gamma=0.0016\omega$ was
taken into account. \textbf{c)} The Higgs-like amplitude mode derived
in Eq. (\ref{eq:higgs_eqs}) is depicted. The numerical solution (blue),
is plotted against the amplitude envelope predicted in Eq. (\ref{eq:higgs_eqs}).
The amplitude trajectory corresponding to the amplitude oscillation
is shown in the inlet of subfigure d) \textbf{d)} Phase profiles of
the amplitude Hamiltonian (\ref{eq:a_b_hamiltonian}). The red curves
indicate the two possible amplitude trajectories $a\left(t\right)$,
$b\left(t\right)$ during the phase transition. The inlet shows the
trajectory corresponding to the damped Higgs-like oscillation of subfigure
b).\label{fig:Fig_collective_modes}}
\end{figure}

These effects are most pronounced in the \textit{weakly damped regime,}
where $h\gg h_{c}$, to which we will stick for the rest of this discussion.
We begin with the soft modes. Note that the phase of the spatial part
of the standing wave in Eq. (\ref{eq:sol_1D}) is chosen spontaneously.
Replacing $x\rightarrow x+\phi/q^{*}$ in Eq. (\ref{eq:sol_1D}),
we obtain another valid solution to Eq. (\ref{eq:PTC_eq}). Following
the usual logic of the Goldstone theorem \citep{altland2010condensed},
we expect that, since a uniform offset $\phi$ is inconsequential
for the system's dynamics, a long-wavelength spatial dependence of
$\phi$ will only have a small impact on the time evolution. It remains
to show that a long-wavelength perturbation of the form $\phi\left(x,t_{0}\right)=\phi_{Q}e^{iQx}$,
where $Q\ll q^{*}$, will propagate through the system as a wave with
frequency $\omega_{G}\left(Q\right)$, and that $\omega_{G}\left(Q\right)\rightarrow0$,
as $Q\rightarrow0$, i.e. the dispersion $\omega_{G}\left(Q\right)$
is soft. Indeed, we find that this is true (see Supplement): For $\left|\phi\right|\ll1$,
the propagation of the Goldstone-like mode is described by the equation
\begin{equation}
\frac{1}{c^{2}}\frac{\partial^{2}\phi}{\partial t^{2}}=\left(1-\frac{1}{3}h\right)\frac{\partial^{2}\phi}{\partial x^{2}}.\label{eq:goldstone_eq}
\end{equation}
The dispersion of the mode is $\omega_{G}\left(Q\right)=c_{G}Q$,
where $c_{G}=\sqrt{1-h/3}c$. Fig. \ref{fig:Fig_collective_modes}a
shows the numerical simulation of Eq. (\ref{eq:PTC_eq}), where a
boundary condition of the form $\phi\left(x=0,t\right)=-0.3\sin\left(\omega_{G}\left(Q\right)t\right)$
was applied. The perturbation at the boundary propagates as a Goldstone-like
mode whose dynamics is described by Eq. (\ref{eq:goldstone_eq}),
and forms a standing wave, such that the phase throughout the PTC
is given by $\phi\left(t,x\right)=\left[\sin\left(\omega_{G}\left(Q\right)t-Qx\right)+\sin\left(\omega_{G}\left(Q\right)t+Qx\right)\right]/2$.
For the simulation, we chose $Q=2\pi/\Lambda$, where $\Lambda=20\lambda$
and $\lambda=2\pi/q^{*}$. We also demonstrate the propagation of
the Goldstone-like mode starting from an initial field $D\left(x\right)=\sqrt{\tilde{h}/\tilde{\beta}}\cos\left(q^{*}x-0.2\sin\left(q^{*}x/5\right)\right)$
at $t=0$. The propagation of the distortion for long times is shown
in Fig. \ref{fig:Fig_collective_modes}b.

In addition to the soft Goldstone mode, the steady-state of Eq. (\ref{eq:sol_1D})
supports spatially uniform oscillations of the amplitudes $a$ and
$b$ around the minima of Eq. (\ref{eq:a_b_minima}). This amplitude
oscillation can be viewed as a massive Higgs mode. To derive the oscillation
frequency, we expand the effective Hamiltonian (\ref{eq:a_b_hamiltonian})
around these minima:
\begin{equation}
H\approx-\frac{\tilde{h}^{2}\Omega}{4\tilde{\beta}}+\tilde{h}\Omega\left(\delta b^{2}+\delta a^{2}\right).\label{eq:Ham_around_min}
\end{equation}
Hamilton's equations read

\begin{align}
\delta\dot{b} & =2\tilde{h}\Omega\delta a\nonumber \\
\delta\dot{a} & =-2\tilde{h}\Omega\delta b.\label{eq:higgs_eqs}
\end{align}
Solving for $a\left(t\right)$, we find $\delta\ddot{a}+\left(2\tilde{h}\Omega\right)^{2}\delta a=0$.
Thus, the PTC supports uniform amplitude oscillations of frequency
$2\tilde{h}\Omega$. For finite $\gamma$, the above equations (\ref{eq:higgs_eqs})
will obtain a damping term and read $\delta\dot{b}=-\gamma b/2+2\tilde{h}\Omega\delta a$,
$\delta\dot{a}=-\gamma a/2-2\tilde{h}\Omega\delta b.$ The solution
of these equations (including damping) is shown in Fig. \ref{fig:Fig_collective_modes}b
together with the field amplitude $D\left(t,2\pi/q^{*}\right)$. We
point out that if no dissipative term is included in Eq. (\ref{eq:PTC_eq}),
the PTC, starting with an initial field distribution, will not converge
to a fixed point. Instead, the amplitudes $a$, $b$ will follow the
Hamiltonian dynamics of Eqs. (\ref{eq:higgs_eqs}). 

\subsection*{Symmetry breaking and pattern formation in two dimensions.}

\begin{figure}
\centering{}\includegraphics[width=1\columnwidth]{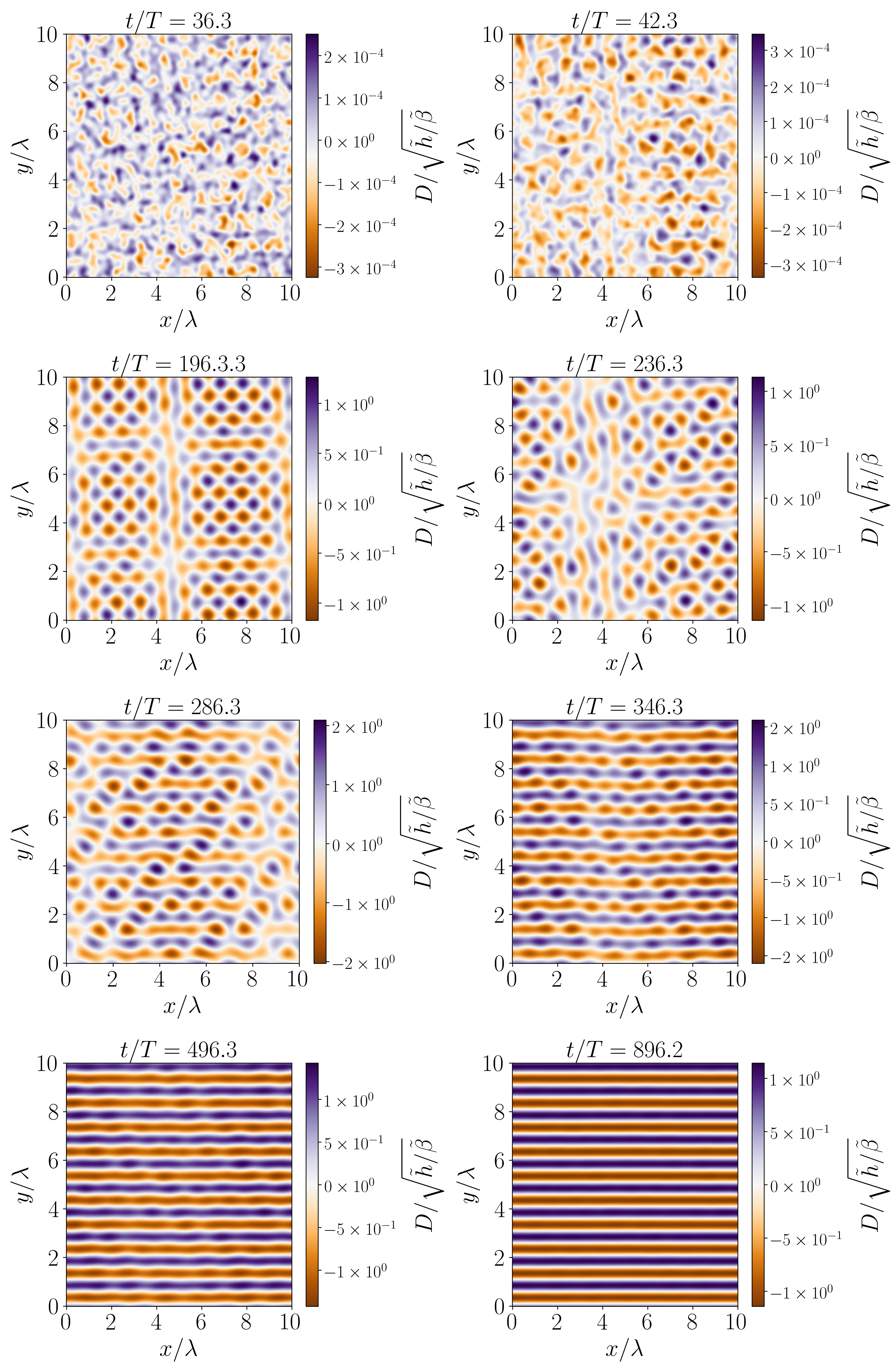}\caption{In two dimensions, nonlinear PTCs evolve to a stripe state that breaks
the translational and rotational symmetries of the system. The figure
shows the simulation of two-dimensional non-linear PTC starting from
random initial conditions on square domain with commensurate periodic
boundary conditions. First, the system develops inhomogeneities whose
length scale is given by the critical $q^{*}$. Finally, after a series
of complex transient patterns, a two dimensional stripe pattern, as
predicted in Eq. (\ref{eq:Hmin_depth}), appears.\label{fig:2d_crystallization}}
\end{figure}
So far we have considered a one-dimensional PTC. A similar symmetry
breaking transition can be observed in higher dimensions. Here, the
spatial second derivative in Eq. (\ref{eq:PTC_eq}) is replaced by
a laplacian: $\partial^{2}D/\partial x^{2}\rightarrow\nabla^{2}D$.
Focusing on the isotropic case, we omit vector signs. The analytical
calculation presented so far can be extended to higher dimensions.
There is, however, a crucial difference to consider. In one spatial
dimension, the only wavenumbers for which the resonance condition
(\ref{eq:resonance condition}) is fulfilled are $\pm q^{*}$. Plane
waves with these two wavenumbers arrange themselfs in a standing wave
pattern as shown in Fig \ref{fig:Fig_transition}c. In two dimensions,
all wavevectors lying on a circle of radius $q^{*}=\Omega/c$ around
the origin are resonant. From studies of Faraday waves, it is known
that parametrically excited waves often arrange themselves in stripes,
square patterns or hexagonal patterns. These arrangements correspond
to one, two or three critical wavevectors $\mathbf{q}^{*}$ that are
chosen spontaneously during the transition to the symmetry breaking
state \citep{muller1994_quasipatterns,chen1999_faraday_amplitude_equations_patterns}.

For the simple nonlinearity of Eq. (\ref{eq:PTC_eq}), we expect a
stripe-pattern \citep{muller1994_quasipatterns} similar to the one-dimensional
scenario. On an infinite domain, the direction of $\mathbf{q}^{*}$
will be chosen spontaneously. To show that the wavepattern for a two
dimensional PTC is indeed a stripe arrangement, we extend the ansatz
(\ref{eq:small_amp_ansatz}) to $N$ wavevectors $\mathbf{q}_{i}^{*}$
with a relative angle $\theta_{N}$, such that $\mathbf{q}_{i}\cdot\mathbf{q}_{i+1}=\mathbf{q}_{N}\cdot\mathbf{q}_{1}=q^{*2}\cos\theta_{N}$
with $\theta_{N}=\pi/N$. The driving term in Eq. (\ref{eq:PTC_eq})
does not supply any momentum to the system. Therefore, each mode with
a wavevector $\mathbf{q}_{i}^{*}$ must be balanced by a mode with
$-\mathbf{q}_{i}^{*}$. Finally, we assume that all modes share the
same time dependent amplitude. Summing up, our ansatz is
\begin{equation}
D=\left(a\left(t\right)\cos\left(\Omega t\right)+b\left(t\right)\sin\left(\Omega t\right)\right)\sum_{i=1}^{N}\cos\left(\mathbf{q}^{*}\cdot\mathbf{x}\right).\label{eq:N_modes_ansatz}
\end{equation}
When inserting this ansatz into Eq. (\ref{eq:PTC_eq}), the linear
terms can be treated similarly as in the 1D case. The nonlinear term,
however, requires additional care. As in the 1D case, we project the
nonlinear part onto the $N$ resonant modes of the ansatz (\ref{eq:N_modes_ansatz})
and ignore fast oscillating, off-resonant terms. The equations of
motion for the amplitudes $a$ and $b$ are similar to Eqs. (\ref{eq:ampl_eoms})
where $\beta$ has to be replaced by $\tilde{\beta}_{N}=9\beta\left(2N-1\right)/32$.
Consequently, we have to make the same replacement $\tilde{\beta}\rightarrow\tilde{\beta}_{N}$
in the potential (\ref{eq:symmetry_breaking_potential}). The new
potential has minima at $a_{N}=-b_{N}=\pm\sqrt{\tilde{h}_{c}/\tilde{\beta}_{N}}\left(2\epsilon\right)^{1/4}$.
The depth of these minima depends on the number of modes $N$:
\begin{equation}
V\left(\bar{b}_{N}=a_{N}-b_{N}\right)\sim\frac{1}{\tilde{\beta}_{N}}\sim\frac{1}{2N-1}.\label{eq:Hmin_depth}
\end{equation}
A similar conclusion can be reached in the weakly damped regime of
Eq. (\ref{eq:a_b_hamiltonian}). The depth of the minimia of the Hamiltonian
again behaves as $\sim1/\tilde{\beta}_{N}$. For more complex nonlinearlities
containing a richer structure of spatial derivatives, the dependence
of the nonlinear term on the angle $\theta$ can compensate for the
tendency of shallower minima for higher $N$. Typically $\theta=\pi/2$,
or $\theta=\pi/3$ is favored, leading to square, or hexagonal wave-lattices.
In general, one expects that the system converges to the deepest minimum
-- in our case, to the $N=1$ minimum, leaving us with a single standing
wave. The orientation and phase are chosen spontaneously. 

Finally, we test our prediction of a stripe-pattern for the field
amplitude in the symmetry broken state by performing a simulation.
Starting from infinitesimal random noise on a square domain of size
$10\lambda$ with periodic boundary conditions, we let the system
evolve in time and track the field amplitude. The results for the
near critical case with $\epsilon=0.3$ are shown in Fig. \ref{fig:2d_crystallization}.
Interestingly, before the PTC reaches the final stripe state, it goes
through a series of nearly-periodic patterns, which can be stable
for many (up to hundreds) oscillation periods.

\section{Discussion}

We investigated the long term behavior of a photonic time crystal
with a generic Kerr nonlinearity. We have shown that the initially
exponential growth of unstable gap modes in these systems heralds
the transition to a new symmetry breaking steady state. In this state,
the field configuration is a standing wave lattice, where the lattice
spacing is determined by the resonant wavenumber $q^{*}$, fulfilling
the resonance condition $\Omega=cq^{*}$. The predicted PTC steady
state exhibits subharmonic response, and breaks the discrete time
translation symmetry of the periodic modulation. This is a common
feature of many types of parametrically driven nonlinear waves \citep{zakharov1975spin_wave_turbulence,edwards1994_faraday_waves},
as it has been pointed out in recent literature on time crystals \citep{yao2017discrete_classical_time_crystals,else2020discrete_time_crystals,zaletel2023colloquium_time_crystals}.
Thus the nonlinear PTC exhibits discrete dissipative time crystalline
behavior in the sense of a phase of matter \citep{zaletel2023colloquium_time_crystals}.

The breaking of continuous spatial translation symmetry is accompanied
by the appearance of soft Goldstone-like modes, which consist of propagating,
wavelike lattice distortions, similar to phonons in a crystal lattice.
In addition, gapped modes that correspond to amplitude oscillations
of the electric field emerge. The dispersions of these emergent modes
depend on the driving strength, which opens the possibility to use
nonlinear PTCs in the symmetry broken state as tunable metamaterials.
We also investigated the behavior of nonlinear PTCs in two spatial
dimensions, and showed that the electric fields self-organize to stripe
patterns that break the translational, and rotational symmetries of
the system. This behavior could help to create controllable, spatially
structured high intensity electric fields. Goldstone-like and gapped
modes as the ones predicted here can be expected to exist in other
parametrically driven, spatially extended classical and quantum systems.
Apart from Faraday waves in classical liquids, where collective behavior
has been observed \citep{domino2016faraday_elastic_metamaterial_goldstone,kucher2023_faraday_collective_goldstone_trains},
important candidates are discrete space-time crystals in trapped BECs
\citep{smits2018observation_space_time_crystal_cold_atoms,liao2019_theory1_space_time_crystal_cold_atoms,smits2020_theory_exp_3_space_time_crystal_cold_atoms,smits2021_theory2_space_time_crystal_cold_atoms,stehouwer2021_theory4_space_time_crystal_cold_atoms}.
\begin{acknowledgments}
We thank S. Gomé and M. Segev for useful discussions. E.K. acknowledges
financial support by the Helen Diller quantum center.
\end{acknowledgments}

\section*{Appendix} 

\renewcommand{\thesection}{Appendix}%

\setcounter{figure}{0}
\renewcommand{\thefigure}{C\ \arabic{figure}}%

\setcounter{equation}{0}
\renewcommand{\theequation}{A\,\arabic{equation}}%

\subsection{Expansion close to the instability threshold. Continuity of the phase
transition.}

Here, we explain the technical details involved in deriving the gradient
descent dynamics of Eq. (\ref{eq:symmetry_breaking_potential}) in
the near-critical regime. First, we reformulate the amplitude equations
(\ref{eq:ampl_eoms}) in terms of the small parameter $\epsilon=\left(h-h_{c}\right)/h_{c}=\left(\tilde{h}-\tilde{h}_{c}\right)/\tilde{h}_{c}$:
\begin{align}
\dot{\bar{a}} & =-\Omega\tilde{h}_{c}\left(2+\epsilon\right)\bar{a}+\frac{\tilde{\beta}\Omega}{2}\bar{b}\left(\bar{a}^{2}+\bar{b}^{2}\right)\nonumber \\
\dot{\bar{b}} & =\Omega\tilde{h}_{c}\epsilon\bar{b}-\frac{\tilde{\beta}\Omega}{2}\bar{a}\left(\bar{a}^{2}+\bar{b}^{2}\right).\label{eq:ampl_eoms-1-1}
\end{align}
The small $\epsilon$ characterizes the growth of the unstable mode,
which defines a slow time scale. We introduce an expansion in terms
of $\epsilon$, such that $\bar{a}=\epsilon^{1/4}\bar{a}_{1}+\epsilon^{3/4}\bar{a}_{2}...$,
$\bar{b}=\epsilon^{1/4}\bar{b}_{1}+\epsilon^{3/4}\bar{b}_{2}+...$
and $\partial_{t}=\partial_{t_{0}}+\epsilon\partial_{t_{1}}$. At
orders $\epsilon^{1/4}$ and $\epsilon^{3/4}$, we obtain
\begin{align*}
\frac{\partial\bar{a}_{1}}{\partial t_{0}} & =-2\Omega\tilde{h}_{c}\bar{a}_{1}\\
\frac{\partial\bar{a}_{2}}{\partial t_{0}} & =-2\Omega\tilde{h}_{c}\bar{a}_{2}-\frac{\tilde{\beta}\Omega}{2}\bar{b}_{1}\left(\bar{a}_{1}^{2}+\bar{b}_{1}^{2}\right).
\end{align*}
These equations describe a quick relaxation towards the fixed point
\begin{align}
\bar{a}_{1} & =0\nonumber \\
\bar{a}_{2} & =-\frac{\tilde{\beta}}{4\tilde{h_{c}}}\bar{b}_{1}^{3}.\label{eq:quick_partial_fixed_point}
\end{align}
At order $\epsilon^{5/4}$, we derive the equation
\[
\frac{\partial\bar{b}_{1}}{\partial t_{1}}=\tilde{h}_{c}\Omega\bar{b}_{1}+\frac{\tilde{\beta}\Omega}{2}\bar{a}_{2}\bar{b}_{1}^{2}.
\]
For a small $\epsilon$, we can assume that the relaxation of $\bar{a}_{2}$
towards the value given in Eq. (\ref{eq:quick_partial_fixed_point})
is infinitely fast, and write
\begin{equation}
\frac{\partial\bar{b}_{1}}{\partial t_{1}}=\tilde{h}_{c}\Omega\bar{b}_{1}-\frac{\tilde{\beta}^{2}\Omega}{8\tilde{h_{c}}}\bar{b}_{1}^{5}.\label{eq:grad_desc_b^5}
\end{equation}
Restoring the original variables $t$ and $\bar{b}$ we obtain
\[
\frac{\partial\bar{b}}{\partial t}=\epsilon\tilde{h}_{c}\Omega\bar{b}-\frac{\tilde{\beta}^{2}\Omega}{8\tilde{h_{c}}}\bar{b}^{5}.
\]
This is the gradient descent dynamics described by Eq. (\ref{eq:symmetry_breaking_potential})
of the main text.

\subsection{Goldstone-like Modes}

\renewcommand{\theequation}{B\,\arabic{equation}}We add a time and
space dependent phase to the steady state solution found in the main
text for low damping:

\begin{equation}
D_{\phi}\left(x,t\right)=\pm\sqrt{\frac{8h}{9\beta}}\sin\left(\Omega t\right)\cos\left(q^{*}x+\phi\left(x,t\right)\right).\label{eq:solution_with_phaseshift}
\end{equation}
Here, $\phi\left(x,t_{0}\right)\ll1$. We seek to derive an equation
governing the dynamics of the phase $\phi\left(x,t\right)$. To this
purpose, we write
\begin{align}
D_{\phi}\left(x,t_{0}\right) & =\frac{A}{2}\sin\left(\Omega t-qx-\phi\left(t,x\right)\right)\nonumber \\
 & \quad+\frac{A}{2}\sin\left(\Omega t+qx+\phi\left(t,x\right)\right),\label{eq:goldstone_ansatz_comb_waves}
\end{align}
where $A=\sqrt{8h/9\beta}$. In what follows, we will use the abbreviations
$\sin\left(\Omega t-qx-\phi\left(t,x\right)\right)=\sin\left(-\right)$
and $\sin\left(\Omega t+qx+\phi\left(t,x\right)\right)=\sin\left(+\right)$
and similar for $\cos$-terms. Inserting Eq. (\ref{eq:goldstone_ansatz_comb_waves})
into Eq. (\ref{eq:PTC_eq}) and comparing the coefficients in front
of the $\cos\left(+\right)$ terms we find an equation for $\phi\left(t,x\right)$:
\begin{equation}
\frac{1}{c^{2}}\frac{\partial^{2}\phi}{\partial t^{2}}=\left(1-\frac{1}{3}h\right)\frac{\partial^{2}\phi}{\partial x^{2}}.\label{eq:goldstone_eq-1}
\end{equation}
The same equation follows from the $\cos\left(-\right)$ terms. The
$\sin\left(\pm\right)$ terms, on the other hand, give corrections
of order $Q/q^{*}$ to the solution (\ref{eq:solution_with_phaseshift}).
These can be neglected for a smooth, small-$Q$ perturbation of the
kind that we are considering. Compared to the velocity of electromagnetic
waves in the undriven medium, the velocity of the Goldstone mode $c_{G}$
is reduced by a factor that depends on the amplitude of the time varying
term in Eq. (\ref{eq:PTC_eq}):
\[
c_{G}=\sqrt{1-\frac{1}{3}h}c.
\]
\bibliographystyle{../thesis-bib}
\bibliography{bibliography}

\end{document}